\documentclass[useAMS,usenatbib]{mn2e}

\usepackage{amsmath}
\usepackage{amssymb}
\usepackage{graphicx}
\usepackage{dcolumn}
\usepackage{bm}
\usepackage{ulem}

\title[Phase diagrams of binary ionic mixtures]
{Phase diagrams of binary ionic mixtures and white dwarf cooling}
\author[D. A. Baiko]{D. A. Baiko\thanks{E-mail:baiko@astro.ioffe.ru} \\
A. F. Ioffe Physical-Technical Institute, 
Politekhnicheskaya 26, 194021 St.-Petersburg, Russia}

\begin{document}

\date{Accepted; Received ; in original form}

\pagerange{\pageref{firstpage}--\pageref{lastpage}} \pubyear{2014}

\maketitle

\label{firstpage}

\begin{abstract}
Phase diagrams of fully ionized binary ionic mixtures are considered
within the framework of the linear mixing formalism taking into 
account recent advances in understanding quantum one-component plasma 
thermodynamics. We have followed a transformation of azeotropic phase 
diagrams into peritectic and eutectic types with increase of the 
charge ratio. For solid $^{12}$C/$^{16}$O and $^{16}$O/$^{20}$Ne 
mixtures, we have found extensive miscibility gaps. Their appearance 
seems to be a robust feature of the theory. The gaps evolve naturally 
into two-solid regions of eutectic phase diagrams at higher $Z_2/Z_1$. 
They do not depend on thermodynamic fit extensions beyond their 
applicability limits. The gaps are sensitive to binary mixture 
composition and physics, being strongly different for C/O and O/Ne 
mixtures and for the three variants of corrections to linear-mixing 
solid-state energies available in the literature. When matter cools to 
its miscibility gap temperature, the exsolution process takes place. 
It results in a separation of heavier and lighter solid solutions. 
This may represent a significant reservoir of gravitational 
energy and should be included in future white dwarf (WD) 
cooling simulations. 
Ion quantum effects mostly resulted in moderate modifications, 
however, for certain $Z_2/Z_1$, these effects can produce qualitative 
restructuring of the phase diagram. This may be important for the 
model with $^{22}$Ne distillation in cooling C/O/Ne WD proposed 
as a solution for the ultramassive WD cooling anomaly.  
\end{abstract}




\section{Introduction}
A good model of matter for such astrophysical objects as 
interior layers of white dwarfs (WD) and outer neutron star crust is 
a fully ionized binary ionic mixture (BIM). This is a system of point 
charges (ions) of two sorts, with different charge and/or mass numbers, 
immersed into a degenerate nearly incompressible electron gas. 
Under realistic conditions, depending on matter density and 
temperature, BIM can be in a liquid or crystallized state. 

Various BIM properties are important for 
astrophysical applications. Since liquid and solidified BIM at 
equilibrium with each other, typically, must have different fractions 
of constituents, ion crystallization in a BIM is accompanied by 
separation of ion species. This leads to gravitational energy release. 
Combined with latent heat release, this effect appears to be crucial for 
WD cooling theory. Recent {\it Gaia} observations 
\citep[][]{G18} indicate a pile-up of WD on the so-called 
Q branch, which implies significant additional cooling delay likely 
associated with other aspects of crystallization physics 
\citep*[e.g.][]{T+19,CCM19,B+20_2,CHC20,C+21,BDS21}.    

Ion structure factor and degree of disorder have an impact on 
electron thermal and electric conduction 
\citep[e.g.][]{FI76,YU80,DG09,RR16,C20,F+20}, which, in turn, determine 
thermal 
profile of compact stars and evolution of their magnetic fields. Ion 
correlations and screening modify thermonuclear reaction 
rates by many orders of magnitude and even allow pycnonuclear burning 
at zero temperature \citep[e.g.][]{YS89,I93}. 
This affects composition of accreted crust in low-mass X-ray binaries 
\citep[e.g.][]{L+18,SC19} as well as regimes and outcomes of helium 
burning in accreting WD \citep*[e.g.][]{PTY14}. 

Thermodynamics of BIM has been the subject of intensive research 
\citep[e.g.][]{SC93,O+93,DWS03,MC10,H+10,B+20}. Due to strong coupling, 
first-principle simulations are preferred to obtain 
reliable information on the properties of the system. Moreover, there 
is an important 
corner of the parameter space, where relatively light 
ions (e.g. helium or carbon) occur at fairly high densities, 
so that ion quantum effects are already pronounced but 
pycnonuclear fusion is not yet rapid enough. In this regime, 
first-principle {\it quantum} simulations are desirable. 

Another approach is based on the idea that BIM thermodynamics can be 
recovered with reasonable accuracy via ``linear mixing'' from 
thermodynamics of appropriate single-species ion systems (known as 
one-component plasmas or OCP). This is certainly less expensive 
than first-principle simulations of BIM. However, besides OCP 
thermodynamics, this approach relies on the knowledge of 
``corrections'' to linear-mixing quantities to make the results more
realistic. Also, it inevitably requires usage of OCP thermodynamic 
fits at temperatures far beyond the aggregate states, for which 
they have been designed, or replacement of the fits by extrapolations 
\citep[e.g.][]{MC10,J+21}.   

In this work, we focus on phase transition properties of BIM within 
this linear mixing with corrections framework. We aim to take 
advantage of the fact that quantum OCP thermodynamics has been
recently developed both in the liquid and in the solid phases
\citep[][]{B19,BY19,BC22}. In particular,
in the solid phase, in addition to the well-known quantum harmonic 
thermodynamics \citep*[e.g.][]{BPY01}, quantum anharmonic functions 
have been calculated and 
parameterized. This allows one, for the first time, to include ion 
quantum effects into BIM phase diagram construction. All previous 
calculations of this sort were strictly classic. 

The paper is organized as follows. In sections \ref{liq} and \ref{sol},
we reproduce the standard expressions for BIM free energies in the 
liquid and solid phases, respectively, paying special attention to 
still existing theoretical uncertainties. In section \ref{eqcond}, we 
formulate equilibrium equations, which determine 
melting/crystallization and solid solubility curves on a phase diagram.
Phase diagrams for C/O and O/Ne mixtures are analyzed in detail in 
sections \ref{UOsec} and \ref{ONesec}, respectively.
Astrophysical implications of miscibility gaps
discovered in C/O and O/Ne systems are discussed in section \ref{astro}. 
In section \ref{varyZ2}, we 
follow a sequence of phase diagrams for systems with intermediate 
charge ratios (including O/Mg, C/Ne, O/Si, and C/Mg mixtures), which 
illustrates a diagram transformation from azeotropic to peritectic and 
finally to eutectic type. Phase diagrams for C/Ne and C/Mg mixtures 
are considered in more detail in section \ref{UNeUMg}, 
where we also make a few remarks on $^{22}$Ne 
distillation process in cooling C/O/Ne WD \citep[][]{BDS21} and on 
eutectic system cooling. 
Section \ref{HeUsec} is devoted to the phase diagram of a He/C 
mixture. We conclude in section \ref{concl}.

\section{Free energy of the liquid phase}
\label{liq}
The Helmholtz free energy of an OCP, 
$F_{\rm OCP}$,  can be written as
\begin{equation}
          F_{\rm OCP} (N,V,T,Z,m) \equiv 
          NT f_{\rm OCP} (N/V,T,Z,m)~, 
\label{Ff}
\end{equation}
where $N$ is the number of ions, $V$ is the volume, $T$ is the 
temperature, $Z$ is the ion charge number, and $m$ is the ion mass. In 
this way, the reduced Helmholtz free energy $f$ (per particle, measured 
in units of $T$) is conveniently 
introduced ($k_{\rm B}\equiv 1$). The same definition 
will be adopted also for various partial contributions to the free 
energy. 

In a liquid phase (`$\ell{\sl iq}$'), it is customary and 
convenient to separate the standard ideal gas contribution (`id') 
and a correction due to the system non-ideality (i.e.\ having to 
do with Coulomb interactions: `nid'):
\begin{eqnarray}
          F^{\ell{\sl iq}}_{\rm OCP}
          &\equiv& NT f^{\ell{\sl iq}}_{\rm OCP}  
\nonumber \\
          &=& F_{\rm id} + F_{\rm nid} \equiv
         NT [f_{\rm id}(\Gamma,\theta) + f_{\rm nid}(\Gamma,\theta)]~.       
\label{Fidnid}
\end{eqnarray}
It is well-known, that the reduced free energy in the OCP model 
depends on just two 
dimensionless combinations of physical parameters: 
$\Gamma= \Gamma(n,T,Z)= Z^2 e^2/(aT)$ and 
$\theta=\theta(n,T,Z,m)=T_{\rm p}/T$. In this case, 
$a=(4\pi n/3)^{-1/3}$ is the ion Wigner-Seitz radius, $n=N/V$ is the 
ion number density, and $T_{\rm p} = \hbar \sqrt{4 \pi n Z^2 e^2/m}$ 
is the ion plasma temperature.    

Furthermore, the non-ideal contribution can be split into classic 
(`cl') and quantum (`q') terms:
\begin{equation}
         F_{\rm nid} = F_{\rm nid\,cl} + F_{\rm nid\,q} \equiv
         NT [f_{\rm nid\,cl}(\Gamma) + f_{\rm nid\,q}(\Gamma,\theta)]~,  
\label{Fclq}
\end{equation}
and the classic term, $f_{\rm nid\,cl}$, is known to be a function of 
just one parameter, $\Gamma$.

Suppose we have a BIM of $N_1$ and $N_2$ ions 
characterized by charge numbers $Z_1$ and $Z_2$ and masses $m_1$ and 
$m_2$, respectively. Throughout the paper, we shall assume that 
$Z_2>Z_1$. We want to come up with a reasonable approximation
for thermodynamic functions of this BIM based on detailed 
knowledge of the OCP thermodynamics. The electron number density in the 
mixture is $n_{\rm e} = (Z_1 N_1 + Z_2 N_2)/V$. Then, in spite of the 
fact that the actual density of the $j$th type ions in the mixture 
is $N_j/V$, $j=1,2$, one is physically motivated to employ non-ideal 
quantities of the $j$th type OCP at ion density $n_{\rm e}/Z_j$, 
which would imply the same electron density in the OCP as in the BIM. 
By contrast, the ideal quantities must depend on true ion number 
densities.   

Thus, according to the standard linear mixing (`LM') 
prescription, for a classic liquid, 
\begin{eqnarray}
       \frac{F^{\ell{\sl iq}}_{\rm BIM\,cl}}{T} &=& 
       \frac{F_{\rm 12\,id}}{T} +
       N_1 f_{\rm 1\,nid\,cl} + N_2 f_{\rm 2\,nid\,cl} 
\nonumber \\
       &+& (N_1+N_2) \Delta f^{\ell{\sl iq}}_{\rm nLM}~. 
\label{FclBIM}
\end{eqnarray}
In this case, $F_{\rm 12\,id}(N_1,N_2,V,T,m_1,m_2)$ is the free energy
of a mixture of two ideal gases with densities $N_1/V$ and $N_2/V$,
$f_{j\,{\rm nid\,cl}} = f_{\rm nid\,cl}(\Gamma_j)$, 
\begin{equation}
  \Gamma_j = \Gamma\left(\frac{n_{\rm e}}{Z_j}, T, Z_j \right) =
  \frac{Z_j^{5/3}e^2}{a_{\rm e}T}~,
\label{Gammaj}
\end{equation}
and $a_{\rm e} = (4 \pi n_{\rm e}/3)^{-1/3}$. Finally, 
$\Delta f^{\ell{\sl iq}}_{\rm nLM}$ is a correction to the linear 
mixing rule in the liquid (`nLM' for `non-linear-mixing'), which can be 
obtained from a first-principle simulation of the mixture or 
from a more advanced theory. 

Let us note, that the use of
$F_{\rm 12\,id}$ in equation (\ref{FclBIM}) ensures that, in the limit 
of high temperatures, the 
entropy of the mixture contains, as it should, mixing entropy of two 
ideal gases with $N_1$ and $N_2$ particles \citep[e.g.][]{LL80}:
\begin{equation}
       S_{\rm mix} = 
       N_1 \ln{\frac{N_1 + N_2}{N_1}}
       + N_2 \ln{\frac{N_1 + N_2}{N_2}}~.
\label{Sid}
\end{equation}

For a quantum liquid, the prescription (\ref{FclBIM}) can be extended
in the natural way:
\begin{equation}
       \frac{F^{\ell{\sl iq}}_{\rm BIM}}{T} = 
       \frac{F_{\rm 12\,id}}{T} +
       N_1 f_{\rm 1\,nid} + N_2 f_{\rm 2\,nid} 
       + (N_1+N_2) \Delta f^{\ell{\sl iq}}_{\rm nLM}~, 
\label{FBIM}
\end{equation}
where $f_{j\,{\rm nid}} = f_{\rm nid}(\Gamma_j,\theta_j)$, and
\begin{equation}
   \theta_j = \theta\left(\frac{n_{\rm e}}{Z_j}, T, Z_j, m_j \right) 
     = \frac{\hbar}{T} \sqrt{\frac{4 \pi n_{\rm e} Z_j e^2}{m_j}}~.
\label{thetaj}
\end{equation}
Equation (\ref{FBIM}) can be rewritten identically as
\begin{eqnarray}
       \frac{F^{\ell{\sl iq}}_{\rm BIM}}{T} &=& 
       N_1 f^{\ell{\sl iq}}_{\rm 1\,OCP} + 
       N_2 f^{\ell{\sl iq}}_{\rm 2\,OCP} - S_{\rm mixZ}
\nonumber \\
       &+& (N_1+N_2) \Delta f^{\ell{\sl iq}}_{\rm nLM}~,
\label{FBIMalt}
\end{eqnarray}
where $f^{\ell{\sl iq}}_{j\,{\rm OCP}} 
= f_{\rm id}(\Gamma_j,\theta_j) + f_{\rm nid}(\Gamma_j,\theta_j)$, and
\begin{equation}
       S_{\rm mixZ} = 
       N_1 \ln{\frac{Z_1 N_1 + Z_2 N_2}{Z_1 N_1}}
       + N_2 \ln{\frac{Z_1 N_1 + Z_2 N_2}{Z_2 N_2}}~.
\label{SZ}
\end{equation}

In practical calculations, we shall take the quantity $f_{\rm nid}$ 
as a sum of classic fit of 
\citet{PC00} and quantum fit of \citet{BC22} at all required
temperatures. Unfortunately, no information on quantum corrections to 
the linear mixing is presently available, thus, we can only use the 
same $\Delta f^{\ell{\sl iq}}_{\rm nLM}$ from classic equation 
(\ref{FclBIM}) in quantum equations (\ref{FBIM}) and (\ref{FBIMalt}).
In what follows, we shall utilize the
fitting formula for $\Delta f^{\ell{\sl iq}}_{\rm nLM}$ 
from \citet{P+09} and shall include this
quantity in all our calculations.

\section{Free energy of the solid phase}
\label{sol}
In a solid phase (`${\sl so}\ell$'), the reduced free energy of the 
OCP reads 
\begin{equation}
       f^{{\sl so}\ell}_{\rm OCP} = f_{\rm Mad}(\Gamma) 
       + f_{\rm h}(\theta) + f_{\rm ah}(\Gamma,\theta)~,
\end{equation}
where indices `Mad', `h', and `ah' stand for Madelung, harmonic, and 
anharmonic contributions, respectively. 

In the linear mixing formalism, in analogy with 
equation (\ref{FBIMalt}), the Helmholtz free energy of a binary solid 
mixture is written as
\begin{eqnarray}
       \frac{F^{{\sl so}\ell}_{\rm BIM}}{T} &=& 
       N_1 f^{{\sl so}\ell}_{\rm 1\,OCP} 
       + N_2 f^{{\sl so}\ell}_{\rm 2\,OCP} - S_{\rm mixZ}
\nonumber \\
       &+& (N_1+N_2) \Delta f^{{\sl so}\ell}_{\rm nLM}~.
\label{Fsol}
\end{eqnarray}
This can be rewritten identically as
\begin{eqnarray}
       \frac{F^{{\sl so}\ell}_{\rm BIM}}{T} &=& 
       \frac{F_{\rm 12\,id}}{T} +
       N_1 (f^{{\sl so}\ell}_{\rm 1\,OCP}-f_{\rm 1\,id}) 
       + N_2 (f^{{\sl so}\ell}_{\rm 2\,OCP}-f_{\rm 2\,id}) 
\nonumber \\
       &+& (N_1+N_2) \Delta f^{{\sl so}\ell}_{\rm nLM}~. 
\label{Fsolalt}
\end{eqnarray}
In this case, 
$f_{j\,{\rm id}} = f_{\rm id}(\Gamma_j,\theta_j)$, while   
$\Delta f^{{\sl so}\ell}_{\rm nLM}$ is a correction to the linear 
mixing prescription in the solid phase, which must be deduced from more 
advanced studies. Let us emphasize once again, that $F_{\rm 12\,id}$ 
employs actual ion number densities, whereas $f_{j\,{\rm id}}$ is
calculated for the ``effective'' $j$th ion density equal to 
$n_{\rm e}/Z_j$. In practical calculations, 
$f^{{\sl so}\ell}_{j\,{\rm OCP}} 
= f^{{\sl so}\ell}_{\rm OCP}(n_{\rm e}/Z_j,T)$
will be given by a sum of the Madelung energy, harmonic quantum fit of 
\citet{BPY01}, and anharmonic quantum fit of \citet{BC22} at all 
required temperatures (but see Fig.\ \ref{cmg_diag}a and section 
\ref{HeUsec}). 

Presently, there exist three different parameterizations of
$\Delta f^{{\sl so}\ell}_{\rm nLM}$. Two of them \citep[][]{O+93,DWS03}
are based on first-principle classic simulations of solid mixtures. 
In reality, they include
only a correction to the Madelung energy. The third one 
\citep[][]{PC13} is an interpolation, attempting to extend the 
results of \citet{DWS03} to higher charge number ratios.

In Fig.\ \ref{dfs_map}a, we compare these parameterizations as functions
of the heavier element fraction, $x_2=N_2/(N_1+N_2)$, for several 
$Z_2/Z_1$ marked 
near the curves. Thick (red), thin (blue), and intermediate (green) 
curves represent 
parameterizations of \citet{O+93}, \citet{DWS03}, and \citet{PC13}, 
respectively. Vertical axis shows $\Delta f$ in units of 
$\Gamma_1 \equiv Z_1^{5/3} \Gamma_{\rm e}$ 
(such normalization is marked by a tilde). It is 
noteworthy, that, for C/O mixture, parameterizations of 
\citet{DWS03} and \citet{PC13} predict noticeably larger 
$\Delta f^{{\sl so}\ell}_{\rm nLM}$ than does the parameterization of 
\citet{O+93} in a wide range of $x_2$. Another 
significant difference is seen in the behaviour of the 
parameterizations at low $x_2$. 

Let us observe, that equation (\ref{Fsol}) seems somewhat artificial,
because, in the solid phase, in view of the melting phase 
transition, there is no requirement to reproduce the
correct mixing entropy of ideal gases at high temperatures. On the 
other hand, at low temperatures, the prescription (\ref{Fsol}) or 
(\ref{Fsolalt}) results in the residual entropy of the solid mixture
at $T=0$ equal to $S_{\rm mixZ}$ ($f^{{\sl so}\ell}_{\rm OCP}$ yields 
zero entropy at $T=0$), justification for which is not clear. For 
instance, the entropy of a random alloy, with ions of two types 
occupying randomly nodes of a perfect crystal lattice, is given by
\begin{eqnarray}
       \ln{\frac{(N_1 + N_2)!}{N_1! N_2!}} &\approx& S_{\rm mix} 
\nonumber \\
       &=&   
       N_1 \ln{\frac{N_1 + N_2}{N_1}}
       + N_2 \ln{\frac{N_1 + N_2}{N_2}}~. 
\label{Sresid}
\end{eqnarray}
Thus, one is tempted to try an alternative expression for
$F^{{\sl so}\ell}_{\rm BIM}$, which is obtained from equation 
(\ref{Fsol}) by replacement $S_{\rm mixZ} \to S_{\rm mix}$. In what 
follows, we shall analyze phase diagram changes stemming from 
such a replacement.

\section{Equilibrium conditions}      
\label{eqcond}
In order to construct a phase diagram, one has to determine physical 
conditions, at which different phases are in thermodynamic 
equilibrium. In view of a very high electron thermal conductivity, we 
assume that the phases are at a fixed temperature $T$. The phases 
must be also at the same pressure. However, the pressure is not 
convenient to work with, because thermodynamic functions, 
such as the energy or the Helmholtz free energy, are calculated 
microscopically at given particle densities. 

The pressure is dominated by the contribution of degenerate
electrons. Hence, the easiest approach is to assume that 
the electron density $n_{\rm e}$ is the same in all phases.
This approach neglects the fact that, in reality, there is 
a jump 
of electron density and pressure at phase boundaries \citep[compensated 
by an ion pressure jump, cf.][]{BC22}, which affects the Helmholtz 
free energies and ion chemical potentials derived from 
them. A more self-consistent approach 
would be to switch to Gibbs free energies. This has been done, e.g., by 
\citet{MC10} within the framework of the perturbation theory. There 
was a difference between results obtained by the two approaches but it 
has not been found to be drastic. \citet{BD21} also independently 
showed (using first-principle simulations) that volume changes during 
the phase transition can be neglected.        

In the present paper, we 
adopt an intermediate approach, where we take 
into account the electron pressure jump only in the leading terms of 
the Helmholtz free energy. 

The chemical potentials of ion species 1 and 2 
can be found in the liquid and solid phases as
\begin{equation}
    \mu_{1,2}^{\ell{\sl iq},{\sl so}\ell} = 
    \left(
    \frac{\partial F^{\ell{\sl iq},{\sl so}\ell}_{\rm BIM}}{\partial 
    N_{1,2}}\right)_{N_{2,1},V,T}~.
\end{equation}
If there is an equilibrium, the chemical potentials of both 
ion species in the liquid and solid phases must be equal: 
\begin{equation}
     \mu_{1}^{\ell{\sl iq}} = \mu_{1}^{{\sl so}\ell}~,  
\label{mu1eq} \\
\end{equation}
\begin{equation}
     \mu_{2}^{\ell{\sl iq}} = \mu_{2}^{{\sl so}\ell}~.  
\label{mu2eq}
\end{equation}
These equations form the basis of the phase diagram construction.

In addition to the ion
contributions to the Helmholtz free energy, there is also an electron 
contribution, which can be approximated by the energy of the 
zero-temperature electron gas:
\begin{equation}
    F_{\rm BIM} \to F_{\rm BIM} + E_{\rm e} = F_{\rm BIM} + 
    \frac{V}{\pi^2} \int_0^{k_{\rm F}(n_{\rm e})} {\rm d}k k^2 
    \varepsilon_k~, 
\nonumber
\end{equation}
where $k_{\rm F}(n_{\rm e}) = (3 \pi^2 n_{\rm e})^{1/3}$ is the 
electron Fermi wavevector, and $\varepsilon_k$ is the electron 
energy at wavevector $k$. Such a correction to $F_{\rm BIM}$ results
in a modification of the ion chemical potentials: 
\begin{eqnarray}
    \mu_{1,2}^{\ell{\sl iq},{\sl so}\ell} &\to& 
    \mu_{1,2}^{\ell{\sl iq},{\sl so}\ell} +
    \Delta \mu_{1,2}^{\ell{\sl iq},{\sl so}\ell} 
\nonumber \\
    &\equiv&
     \mu_{1,2}^{\ell{\sl iq},{\sl so}\ell} + 
     Z_{1,2} \varepsilon_{k_{\rm F}} 
     (n_{\rm e}^{\ell{\sl iq},{\sl so}\ell})~,
\end{eqnarray}
where $\varepsilon_{k_{\rm F}}$ is the electron Fermi energy, and we 
have stressed its dependence on $n_{\rm e}$.

Making such an adjustment to the chemical potentials in equations
(\ref{mu1eq}) and (\ref{mu2eq}) (divided by $Z_1$ and $Z_2$, 
respectively), and subtracting the second of them from the first one,
we obtain an equilibrium condition in the form
\begin{equation}
        \frac{\mu_{1}^{\ell{\sl iq}} - \mu_{1}^{{\sl so}\ell}}{Z_1} 
        - \frac{\mu_{2}^{\ell{\sl iq}} - \mu_{2}^{{\sl so}\ell}}{Z_2} 
        = 0~. 
\label{eq1}
\end{equation}
In this way, the electron Fermi energy, which is large and slightly 
different in the liquid and solid phases, cancels out.

The other equilibrium condition can be obtained if one expresses 
the electron pressure increment due to the electron density jump 
($\delta n_{\rm e} = n_{\rm e}^{\ell{\sl iq}}-n_{\rm e}^{{\sl so}\ell}$) 
via that of the ion chemical potential. 
In particular,
\begin{equation}
        P_{\rm e} = -\frac{\partial E_{\rm e}}{\partial V} = 
        -\frac{E_{\rm e}}{V} + n_{\rm e} \varepsilon_{k_{\rm F}}~. 
\end{equation}
Hence,
\begin{equation}
        \frac{{\rm d} P_{\rm e}}{{\rm d} n_{\rm e}}  \delta n_{\rm e} = 
        n_{\rm e} \frac{{\rm d} 
        \varepsilon_{k_{\rm F}}}{{\rm d} n_{\rm e}} \delta n_{\rm e}~. 
\end{equation}
On the other hand,
\begin{equation}
        \frac{{\rm d} \Delta \mu_{1,2}}{{\rm d} n_{\rm e}} 
        \delta n_{\rm e} = 
        Z_{1,2} \frac{{\rm d} 
        \varepsilon_{k_{\rm F}}}{{\rm d} n_{\rm e}} \delta n_{\rm e}~. 
\end{equation}
Thus, subtracting the pressure equality condition 
(divided by $n_{\rm e}$) from the equality of the chemical potentials of 
the first ion species (divided by $Z_1$), we arrive at the second 
equilibrium condition, in which the leading effect of the electron 
density jump is also removed:
\begin{equation}
        \mu_{1}^{\ell{\sl iq}} - \mu_{1}^{{\sl so}\ell} 
        - \frac{P_{\rm BIM}^{\ell{\sl iq}} 
        - P_{\rm BIM}^{{\sl so}\ell}}{n_{\rm e}/Z_1} = 0~. 
\label{eq2}
\end{equation}
In this case, $P_{\rm BIM}$ is the ion pressure in the mixture obtained 
by an appropriate differentiation of equation (\ref{FBIM}) or 
(\ref{Fsol}). All the quantities appearing in equations (\ref{eq1}) and
(\ref{eq2}) should be calculated at one and the same electron density 
$n_{\rm e}$. 

\begin{figure*}
\begin{center}
\leavevmode
\includegraphics[bb=71 548 568 741, width=170mm]{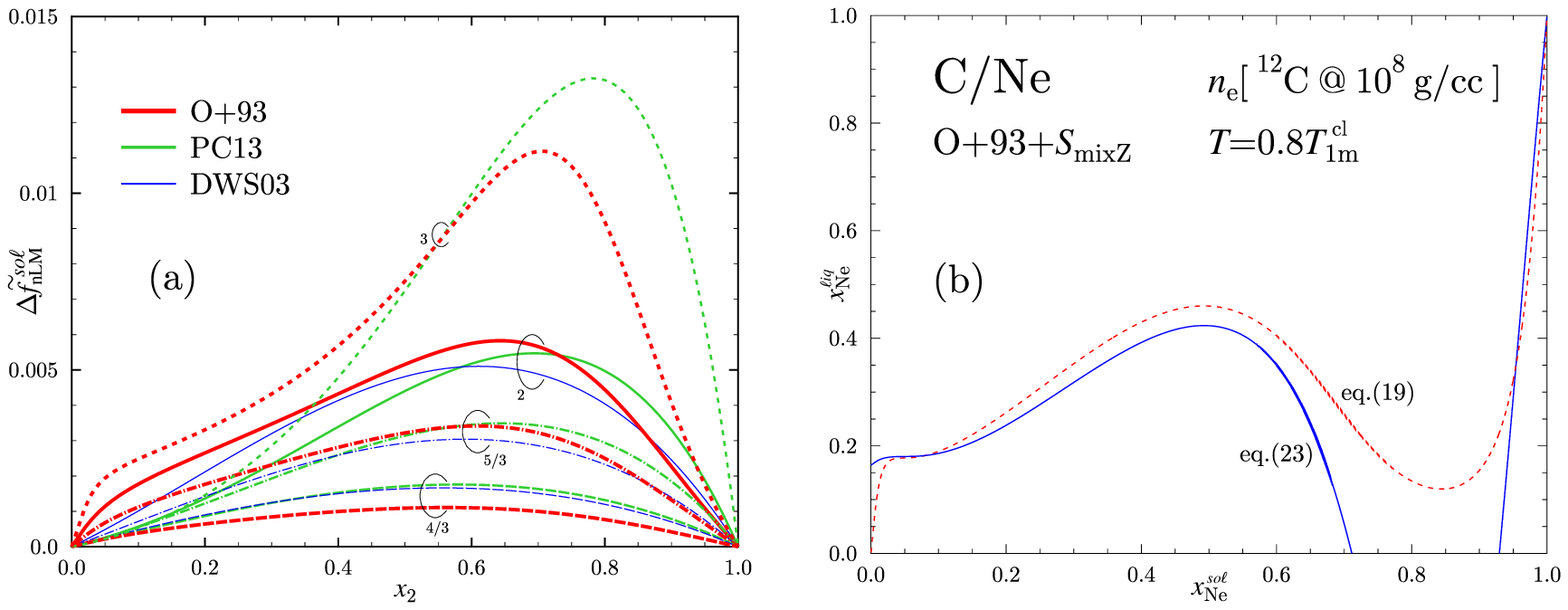}
\end{center}
\vspace{-0.4cm}
\caption[ ]{(a) Various parameterizations of 
$\Delta f^{{\sl so}\ell}_{\rm nLM}$ (normalized to $\Gamma_1$). 
Thick (red), thin (blue), and intermediate (green) curves show results
of \citet{O+93}, \citet{DWS03}, and \citet{PC13}, respectively. Dashed, 
dot-dashed, solid, and short-dashed curves correspond to $Z_2/Z_1 = 4/3$,
$5/3$, 2, and 3, respectively; (b) an example of simultaneous 
solution of equations (\ref{eq1}) (dashed, red) and (\ref{eq2}) 
(solid, blue) for an almost classic $^{12}$C/$^{20}$Ne mixture 
(see text for details).}
\label{dfs_map}
\end{figure*}

In Fig.\ \ref{dfs_map}b, we illustrate conditions (\ref{eq1}) and 
(\ref{eq2}) by dashed (red) and solid (blue) lines, respectively. 
In particular, we 
consider an almost classic $^{12}$C/$^{20}$Ne mixture at 
$n_{\rm e}$, corresponding to pure carbon at $10^8$ g cm$^{-3}$, and 
at $T=0.8 T_{\rm 1m}^{\rm cl}$. This density is relatively low, so 
that ion 
quantum effects for carbon and neon are not yet pronounced. 
$T_{\rm 1m}^{\rm cl} = Z_1^{5/3} e^2/(a_{\rm e} \Gamma_{\rm m})$ is the 
melting temperature of the OCP, composed of 
the first ion species (i.e. carbon here), at given $n_{\rm e}$ and 
assuming classic Coulomb 
coupling strength at melting, $\Gamma_{\rm m} = 175.6$
\citep[][]{BC22}.   
Furthermore, we used $S_{\rm mixZ}$ in equation 
(\ref{Fsol}) and $\Delta f^{{\sl so}\ell}_{\rm nLM}$ parameterized as 
in \citet{O+93}. 

Horizontal and vertical axes display neon fraction 
$x_{\rm Ne} = N_{\rm Ne}/(N_{\rm Ne}+N_{\rm C})$ in the solid and in 
the liquid, respectively. One observes 
two intersection points (there is no intersection in the upper right 
corner), where both conditions are fulfilled simultaneously. 
For specified $n_{\rm e}$ and $T$, these 
points represent  
neon fractions, at which liquid and solid 
phases can co-exist. The locus of such points, obtained as 
$T$ is varied, produces the phase diagram of the binary mixture at 
given density (cf. Fig.\ \ref{cne_diag}a).   

\begin{figure}
\begin{center}
\leavevmode
\includegraphics[bb=15 4 729 556, width=84mm]{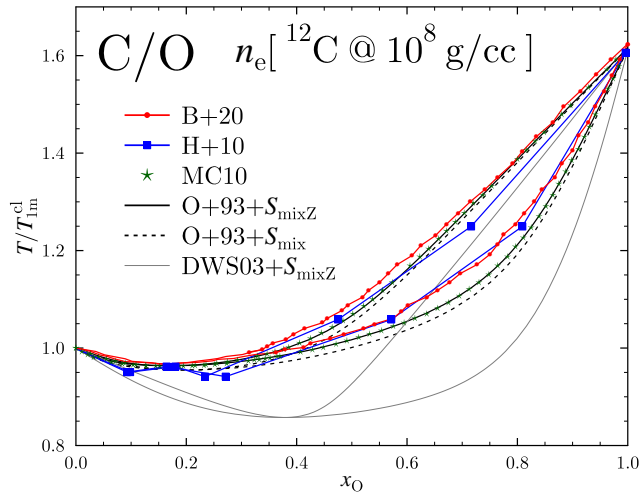}
\end{center}
\vspace{-0.4cm}
\caption[ ]{C/O phase diagram in the temperature-oxygen fraction plane
at almost classic electron density corresponding to pure carbon at 
$10^8$ g cm$^{-3}$. Curves and symbols represent different approaches
to the diagram construction as detailed in the text.}
\label{co_compar}
\end{figure}

Another important application of equations (\ref{eq1}) and (\ref{eq2}) 
is in finding boundaries of unstable solid domains, where a mechanical 
mixture of two 
solid solutions with different fractions of the constituents
is preferable to a single solid solution. In order to accomplish
this, one has to replace all quantities pertaining to the liquid in 
equations (\ref{eq1}) and (\ref{eq2}) by another set of solid 
quantities, parameterized by the second, independent heavier ion 
fraction $x^{{\sl so}\ell}_2$ (which would then appear in place of 
$x^{\ell{\sl iq}}_{\rm Ne}$ in Fig.\ \ref{dfs_map}b). 
These solid solubility boundaries extend the phase diagram of a 
mixture to lower temperatures. At intermediate temperatures, they may 
(but do not have to) connect with melting/crystallization curves found 
at the previous step. By construction, in the present linear 
mixing formalism, one cannot establish dependence of the solid 
solubility boundaries on ion quantum 
effects. For that, one would need to know, in what way quantum effects 
modify $\Delta f^{{\sl so}\ell}_{\rm nLM}$, which would require new 
quantum first-principle studies. 

Before proceeding to the results, let us remark, that equations 
(\ref{eq1}) and (\ref{eq2}) allow one to construct all possible 
double tangents to solid and liquid (or solid and solid) free energy 
curves \citep[cf.][]{G68}, including those
that do not have real thermodynamic significance, i.e. do not correspond
to free energy minima. When plotting phase diagrams, we have checked, 
that selected solutions did in fact minimized the free energy.

\section{Phase diagrams}
\subsection{$^{12}$C/$^{16}$O mixture}
\label{UOsec}
\subsubsection{State of the art}
Let us apply the methods described in sections 
\ref{liq}--\ref{eqcond} to several specific mixtures 
anticipated in the interiors of degenerate stars. Of paramount 
importance is the carbon/oxygen mixture, which is believed to reside 
in the cores of a great number of intermediate-mass WD.    

In Fig.\ \ref{co_compar}, we illustrate state of the art.
Curves and symbols of different types show phase diagrams obtained by 
different authors. The electron density is that of pure carbon at 
$10^8$ g cm$^{-3}$, which is sufficiently low to render ion quantum 
effects not important. The temperature is measured in units of 
$T_{\rm 1m}^{\rm cl}$, which (as before) is the classic melting 
temperature of pure carbon at this electron density.

There are several domains of interest in the temperature-oxygen 
fraction plane. The high-temperature region corresponds 
to a stable liquid mixture, the curve bounding it from below is 
called {\it liquidus}. In the same way, the low-temperature region 
represents a stable solid mixture, which is 
limited from above by the curve called {\it solidus}. Horizontal 
lines at intermediate temperatures cross the phase diagram at one or 
two pairs of points. These specify oxygen fractions in the liquid and 
in the solid, at which the phases can co-exist. The areas 
between the liquidus and solidus describe unstable zones, where matter 
can be viewed as a mechanical mixture of solid and liquid phases, whose 
compositions at a chosen temperature are given by the intersection 
points of solidus or liquidus with the respective horizontal line 
\citep[cf.][]{G68}.

Filled (red) circles display the most recent (classic) results for the 
C/O phase 
diagram obtained in \citet{B+20} using the Gibbs-Duhem integration 
technique coupled to Monte Carlo simulations (GDMC). Slightly older 
molecular dynamics results by \citet{H+10} are shown by (blue) 
squares. Stars (green) show the results of \citet{MC10} obtained 
within the linear mixing with corrections framework as reproduced in 
\citet{B+20}. \citet{MC10} used classic OCP thermodynamics, 
corrections to linear mixing in the solid as fitted by \citet{O+93}, 
double-tangent construction, thermodynamic extrapolation,  
and $S_{\rm mixZ}$ as the residual entropy of the solid mixture. 

The results of \citet{MC10} (stars) are in a perfect agreement with our 
solid (black) line. The latter is obtained under the same 
assumptions except that 
({\it i}) the equilibrium conditions (\ref{eq1}) and (\ref{eq2}) are 
used in place of the double-tangent construction and 
({\it ii}) the fitting expressions for the OCP thermodynamics 
(from \citealt{BPY01,PC00}, and \citealt{BC22} for harmonic solid, 
classic liquid, and quantum liquid and anharmonic solid terms, 
respectively) are utilized directly in the entire temperature range 
without invoking any extrapolation procedures. Dashes differ from the 
solid line by the replacement $S_{\rm mixZ} \to S_{\rm mix}$ in 
equation (\ref{Fsol}). The difference with the solid line 
is seen to be moderate, whereas the 
disagreement with the results of \citet{B+20} becomes worse. Finally, 
the thin solid (grey) curve is obtained by the same method as the 
solid (black) one except that the parameterization of 
$\Delta f^{{\sl so}\ell}_{\rm nLM}$ is taken from \citet{DWS03}. It is 
evident, that this result is in a more serious disagreement with the 
other curves than their mutual disagreement.       
 
\begin{figure*}
\begin{center}
\leavevmode
\includegraphics[bb=71 558 569 741, width=170mm]{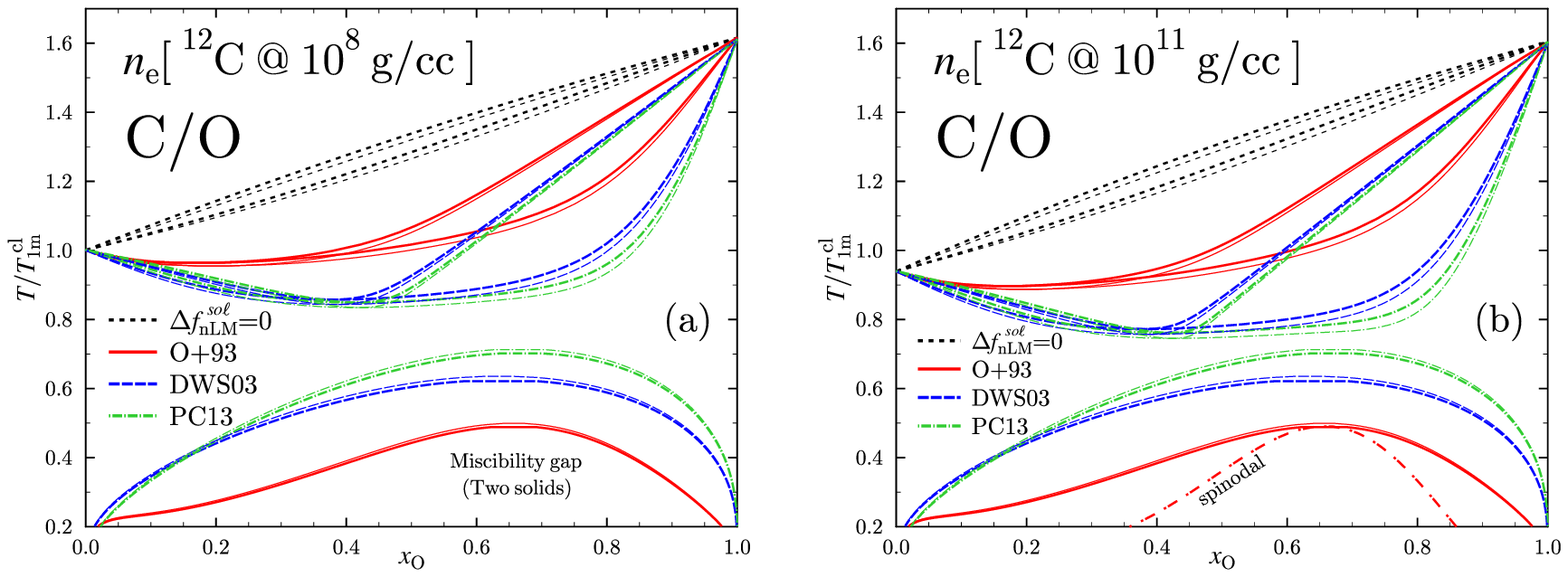}
\end{center}
\vspace{-0.4cm}
\caption[ ]{C/O phase diagram calculated in this work using different 
fits for $\Delta f^{{\sl so}\ell}_{\rm nLM}$: 
$\Delta f^{{\sl so}\ell}_{\rm nLM}$=0 (short-dashed, black), 
\citet{O+93} (solid, red), \citet{DWS03} (long-dashed, blue), and 
\citet{PC13} (dot-dashed, green); different solid mixture residual 
entropies: $S_{\rm mixZ}$ (thick) and $S_{\rm mix}$ (thin); and
different electron densities: almost classic (a) and moderately 
quantum (b). Dot-dash-spaced line shows spinodal for 
O+93+S$_{\rm mixZ}$, same for both densities.}
\label{co_diag}
\end{figure*}

\subsubsection{Miscibility gap and quantum effects}
In Fig.\ \ref{co_diag}, we extend the consideration 
to lower temperatures and higher densities to investigate the 
importance of ion quantum effects. The electron 
density in panel (a) corresponds to pure carbon at 
$10^8$ g cm$^{-3}$, which is three orders of magnitude smaller than 
the density in panel (b). We do not consider densities higher than 
$10^{11}$ g cm$^{-3}$, because 
this would take us beyond the ``stability strip'', and carbon would 
rapidly burn \citep[e.g.][]{B21}. 
Thick and thin curves assume 
the residual entropy of the solid equal to $S_{\rm mixZ}$ and 
$S_{\rm mix}$, respectively. Different line types (and colors) 
refer to different
parameterizations of $\Delta f^{{\sl so}\ell}_{\rm nLM}$: short-dashed 
(black) is for
$\Delta f^{{\sl so}\ell}_{\rm nLM} =0$, while solid (red), long-dashed 
(blue), and dot-dashed (green) 
represent parameterizations proposed by \citet{O+93,DWS03}, and 
\citet{PC13}, respectively.

Regarding the already familiar liquid-solid structure (discussed in 
connection with Fig.\ \ref{co_compar}), several further conclusions can 
be made. While 
for $\Delta f^{{\sl so}\ell}_{\rm nLM}=0$ this part of the diagram 
is of the spindle type, the 
three non-zero variants of $\Delta f^{{\sl so}\ell}_{\rm nLM}$ produce
azeotropic structures. The diagram based on 
$\Delta f^{{\sl so}\ell}_{\rm nLM}$ due to \citet{O+93} is much closer 
to the results of GDMC studies (cf.\ Fig.\ \ref{co_compar}) being 
qualitatively
different from the diagrams based on $\Delta f^{{\sl so}\ell}_{\rm nLM}$
due to the other authors. The difference between predictions based
on fits of \citet{DWS03} and \citet{PC13} is moderate. Use of 
$S_{\rm mix}$ in place of $S_{\rm mixZ}$ results in a moderate shift 
of the curves to lower temperatures 
with their endpoints fixed at OCP melting temperatures. Ion 
quantum effects for this mixture are seen to result in a stretch of 
the liquid-solid structure associated with the fact that, for 
pure carbon 
(left-hand edge of the diagram), quantum effects at the chosen density 
are already sufficiently pronounced to reduce the melting temperature
below its classic value. Liquid-solid 
unstable domains for parameterizations of \citet{DWS03} and 
\citet{PC13} cover noticeably more area at higher density (which means 
stronger 
separation of species upon crystallization), while dot-dashed (green) 
solidus for $0.4 \lesssim x_{\rm O} \lesssim 0.8$ is somewhat flatter.     

In addition to the liquid-solid structure, a 
lower-temperature solid-solid structure exists. Instead of a single 
stable solid solution, we actually obtain a rather extensive unstable 
domain, where a mechanical mixture of two solids is thermodynamically 
preferable. Such regions are well-known in many terrestrial materials 
\citep[e.g. section 4.14 in][]{G68} and are called 
{\it miscibility gaps} (MG). The curve bounding a MG is called 
{\it solvus}. The compositions of the two solid solutions at a chosen 
temperature are given by the intersection points of the solvus with 
the respective horizontal line.  

Clearly, solvus is very sensitive to a particular expression for
$\Delta f^{{\sl so}\ell}_{\rm nLM}$, and the fits of \citet{DWS03} and 
\citet{PC13} generate really large MG. By contrast, the residual 
entropy has almost no effect on MG. The solvi shown in panels (a) and 
(b) are the same, because, as mentioned in section \ref{eqcond}, one 
cannot establish MG dependence on ion quantum effects in the present 
formalism.

\subsection{$^{16}$O/$^{20}$Ne mixture}
\label{ONesec}
\begin{figure*}
\begin{center}
\leavevmode
\includegraphics[bb=71 557 569 741, width=170mm]{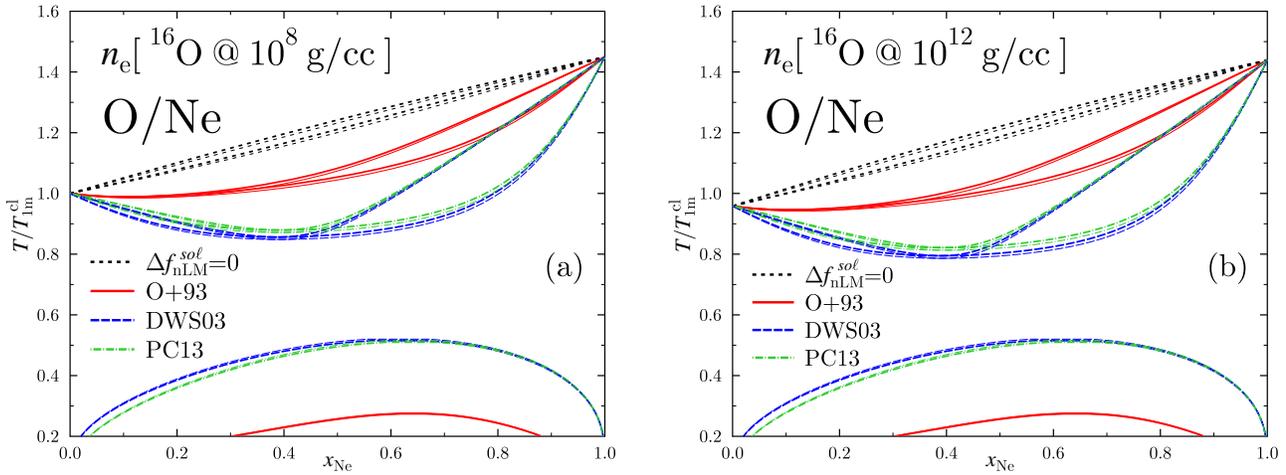}
\end{center}
\vspace{-0.4cm}
\caption[ ]{Same as in Fig.\ \ref{co_diag} but for O/Ne mixture, 
and the electron density in panels (a) and (b) corresponds to pure 
oxygen at $10^8$ and $10^{12}$ g cm$^{-3}$, respectively.}
\label{one_diag}
\end{figure*}

The oxygen/neon mixture is believed to occur in the cores of massive 
WD. In this case, the ratio $Z_2/Z_1=5/4$ is closer to 1 
than for the C/O mixture, hence, one anticipates a 
qualitatively similar but less ``developed'' picture as for the 
latter. In Fig.\ \ref{one_diag}, 
we plot phase diagrams for this mixture. Panel (a) corresponds to the
same electron density as in Fig.\ \ref{co_diag}a for the 
C/O mixture,
whereas quantum effects in panel (b) are illustrated at an order of 
magnitude higher electron density (for the sake of the illustration, we 
neglect neutronization and any other effects which may render such 
physical conditions unachievable). In view of the smaller charge 
ratio, the range of temperatures, at which liquid and solid phases
with different fractions of constituents co-exist, is smaller than for 
C/O. Quantum effects in panel (b) are less pronounced for heavier and 
stronger charged oxygen ions than for carbon ions in 
Fig.\ \ref{co_diag}b. 
This is because the ion plasma temperature scales as $\sqrt{n_{\rm e}}$, 
i.e. for oxygen at $10^{12}$ g cm$^{-3}$ it is $\sim 3.2$ times 
greater than for carbon at $10^{11}$ g cm$^{-3}$. However, the melting 
temperature behaves as $Z_1^{5/3} n_{\rm e}^{1/3}$, and it is 
$\sim 3.5$ times greater for oxygen than for carbon. Accordingly, 
$T/T_{\rm p}$ is greater in Fig.\ \ref{one_diag}b, and this system is 
less quantum.     

Even though $Z_2/Z_1=5/4=15/12$ is not much smaller than $4/3=16/12$,
the effect on MG is profound. Regardless of the 
$\Delta f^{{\sl so}\ell}_{\rm nLM}$ fit, MG for O/Ne are greatly 
reduced in comparison with C/O. 
In particular, the maximum gap temperature for the fit
of \citet{O+93} is just 0.27$T_{\rm 1m}^{\rm cl}$ for O/Ne, whereas
for C/O it reaches 0.49$T_{\rm 1m}^{\rm cl}$.

\subsection{Miscibility gaps and white dwarf cooling}
\label{astro}
The occurrence and properties of the miscibility gaps in Figs.\ 
\ref{co_diag} and \ref{one_diag} may have important implications for 
WD cooling theory. Let us note first, that the appearance of MG is 
rather robust. With increase of $Z_2/Z_1$, they grow and eventually 
collide with the liquid-solid structure, morphing, in a natural way, 
into two-solid regions of eutectic phase diagrams prevailing there 
(cf. Fig.\ \ref{shapes}). Presence of MG does not rely on 
thermodynamic fit extrapolations, as both components are below their 
solidification temperatures, and only formulae for solids are used. 
The gaps are very sensitive to composition and constitutive physics, 
being significantly different for C/O and O/Ne mixtures and for 
different parameterizations of $\Delta f^{{\sl so}\ell}_{\rm nLM}$
(cf.\ Figs.\ \ref{co_diag} and \ref{one_diag}). 

Cooling through MG is similar to cooling through the usual liquid-solid
two-phase region \citep[e.g. section 4.15 in][]{G68}. On reaching the 
solvus curve, a single stable solid solution, prevailing at higher 
temperatures, becomes saturated with a second solid solution, whose 
composition is determined by the other intersection of the solvus with 
the horizontal (constant temperature) line, and exsolution takes 
place. Upon further cooling, the two solutions move down along the 
solvus branches, their relative fractions being determined by the 
lever rule. Such exsolution process in a solid phase is well-known on 
Earth, where it can result in a formation of microscopic to megascopic 
lamellae.\footnote{see e.g. megascopic antiperthite here: 
https://en.wikipedia.org/wiki/Perthite}

Whether a separation of $Z_1$- and $Z_2$-rich solutions does happen in 
dense crystallized interior of compact stars is an open question. In 
principle, Coulomb crystals are in many respects similar to 
strongly-coupled Coulomb liquids, for which such 
a rearrangement would not raise any objections. In both cases, the 
stress tensor is dominated by the pressure of degenerate electron gas, 
whose energy density is huge in comparison with any Coulomb 
contribution responsible for structure. The jumps of thermodynamic 
and kinetic properties at crystallization are not large 
\citep[e.g.][]{B+98,BC22} and eigenmodes of Coulomb solids and liquids 
are alike \citep[e.g.][]{O+13}. Thus, it almost looks as if the 
separation of species with a formation of some geometric (e.g. lamellar) 
superstructure in these peculiar crystals in dense matter should 
proceed easier (with lower barriers) than in terrestrial materials.  
Just like on Earth, the transformation could be intensified by a 
finite amount of undercooling, which helps overcome size and 
compositional barriers. This would make the effective gaps more narrow,  
which can be described as a replacement of the equilibrium solvus 
curves in Figs.\ \ref{co_diag} and \ref{one_diag} by spinodal ones   
\citep[e.g. section 4.15 in][]{G68}.

Turning back to WD cooling, if the lamellae have a tendency of 
forming perpendicular to the gravity, then, in a slab of stellar 
material, we expect accumulation of the heavier ($Z_2$-rich) solution 
at the bottom. This would lead to an energy release. For megascopic 
lamellae, the released energy may be large, because the process 
involves the bulk of the core material, or, at least, those regions, 
where the initial solid composition is such that it hits the MG with 
decrease of temperature. This last consideration also governs 
selectivity of the phenomenon: if, for instance, the composition after 
crystallization in a C/O WD lies beyond the spinodal curve in 
Fig.\ \ref{co_diag}, the exsolution would not operate as effectively
or at all.
 
Naturally, the exsolution in a chunk of stellar material may occur only 
after its crystallization is complete. As we have already mentioned, 
the proximity of MG to the solidus is subject to theoretical 
uncertainty and is sensitively regulated by matter composition.
For instance,  the situation is very different in C/O and O/Ne WD, 
because the gaps occur at significantly higher temperatures in the 
former. The maximum gap temperature may actually be closer to the 
crystallization region, because the simplest linear-mixing 
prescription of section \ref{sol} may underestimate the actual gaps 
due to possible formation of compoundlike structures 
\citep[cf. Fig.\ 6.2 in][]{G68} known as binary ionic crystals 
\citep[e.g.][]{KB15}. Furthermore, ion quantum effects tend to bring 
solidus closer to the solvus at high densities. Given composition, 
the exsolution starts operating earlier in more massive stars, which 
reach relevant temperatures faster. Under certain assumptions, the 
exsolution in central parts of the core may be concurrent with 
solidification of its outer regions.

Summarizing, the MG in C/O and O/Ne mixtures may be responsible for 
substantial reservoirs of gravitational energy in cooling WD. We 
encourage WD modellers to include MG in future cooling simulations 
under various assumptions regarding maximum gap temperature, width, 
and lamella geometry in order to see if any of these parameters can be 
constrained experimentally.

\subsection{Varying $Z_2$} 
\label{varyZ2}
In this subsection, we analyze mixtures, which are 
characterized by a gradually increasing charge ratio. This allows one 
to investigate the evolution of the binary ionic mixture phase diagram 
with this parameter (regardless of the practical relevance of 
considered compositions). In all cases, we assume $Z_1=6$, and mass 
numbers equal to double charge numbers ($A_{1,2}=2Z_{1,2}$).

\begin{figure*}
\begin{center}
\leavevmode
\includegraphics[bb=0 22 1290 983, width=170mm]{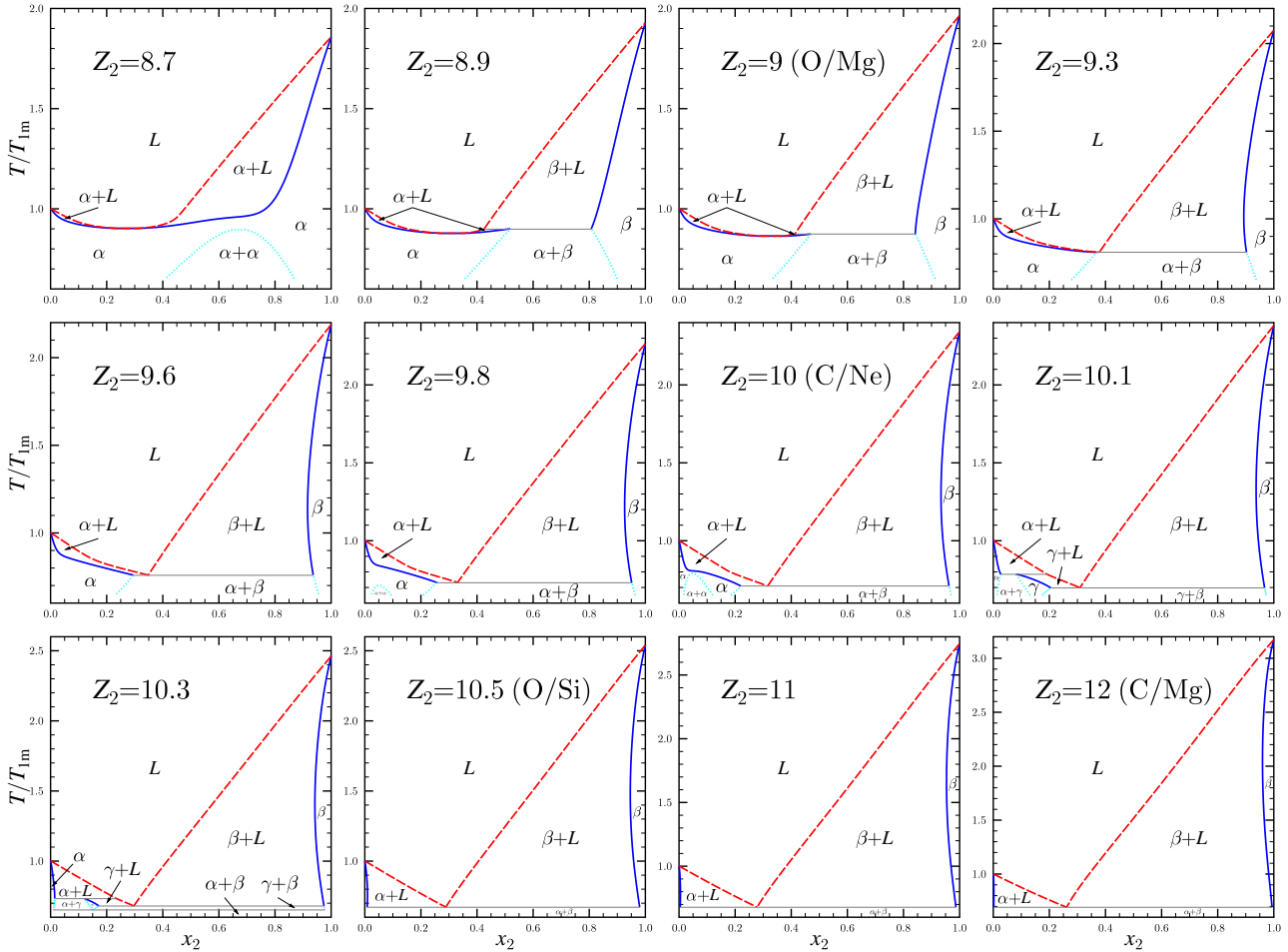}
\end{center}
\vspace{-0.4cm}
\caption[ ]{Phase diagrams for classic mixtures with $Z_1=6$, varying 
$Z_2$, and $A_{1,2}=2Z_{1,2}$. Solidus and liquidus are shown by 
solid (blue) and dashed (red) curves, respectively. Dotted (cyan) 
curves outline solid solubility limits. `{\it L}' marks 
liquid domains; `$\alpha$', `$\beta$', and `$\gamma$' represent 
three different solid solutions. `$+$' sign indicates unstable zones, 
where a mechanical mixture of two solutions is preferable. Thin 
horizontal lines correspond to invariant regions in peritectic
and eutectic systems. 
}
\label{shapes}
\end{figure*}

In Fig.\ \ref{shapes}, there are 12 panels, corresponding to mixtures
with $Z_2$ varying from 8.7 to 12. In all cases, the electron density 
is that of pure carbon at $\rho=10^8$ g cm$^{-3}$, so that the systems 
are essentially classic. Parameterization of 
$\Delta f^{{\sl so}\ell}_{\rm nLM}$ from \citet{O+93} is used. 
Residual entropy of the solid mixture is taken as $S_{\rm mixZ}$. 
Horizontal axes display the fractions of respective $Z_2$-constituents, 
$x_2=N_2/(N_1+N_2)$, vertical axes show temperature in units of 
$T_{\rm 1m}^{\rm cl}$. 

Phase diagrams are comprised of curves of three 
different types: solid (blue), dashed (red), and dotted 
(cyan). Dashed and solid curves determine boundaries of liquid and 
crystal phases, i.e. these are liquidus and solidus, respectively. 
Dotted curves are obtained by solving equations 
(\ref{eq1}) and (\ref{eq2}) applied to de-mixing in the solid phase, 
i.e. they represent solvi.

A liquid mixture with a particular composition and temperature, 
corresponding to a point on a liquidus, can co-exist with only one 
solid mixture, corresponding to a point at the same temperature on the 
associated solidus (cf.\ Fig.\ \ref{dfs_map}b). If two or more pairs 
of solid and dashed curves exist at a given temperature, one can 
establish the association between them based on continuity, 
smoothness, and the fact that associated curves span the same 
temperature range. Likewise, solvus can be split into pairs of 
associated branches, and a point on a particular solvus branch 
represents a solid mixture, which can co-exist with another solid 
mixture with a different composition given by a point at the same 
temperature on the associated solvus branch.
 
In the first panel of Fig.\ \ref{shapes}, $Z_2 = 8.7$, one
sees an azeotropic phase diagram with the MG reaching fairly high 
temperatures, which has developed in a straightforward manner from the 
thick solid (red) line in Fig.\ \ref{co_diag}a (for $Z_2=8$). 
At $Z_2 = 8.9$, the high- and low-temperature structures have 
already merged. The diagram exhibits a $\Lambda$-like feature 
(at $x_2 \approx 0.52$) and 
an invariant reaction of the peritectic type: $L+\beta \to \alpha$. 
This state continues past $Z_2=9$ and includes O/Mg mixture. By 
$Z_2 \approx 9.3$, the diagram becomes eutectic (the 
eutectic point is given by the intersection of the two dashed curves 
or by the tip of the V-feature at $x_2 \approx 0.38$) with the 
standard eutectic invariant 
reaction $L \to \alpha+\beta$ and with a rather large allowed fraction 
of the heavier element in a crystal made of the lighter species 
($x_2 \lesssim 0.36$).

This state continues till $Z_2 \sim 9.8$, where a second MG appears 
at low $x_2$. For C/Ne, the MG barely clears the liquid-solid structure,
crashing into it at $Z_2 \sim 10.1$. For a narrow range of charge 
ratios, this makes the diagram very complex with as many as three
invariant regions\footnote{for binary systems, invariant regions are 
phase diagram regions, where the number of different phases co-existing 
in equilibrium is {\it three}} (for peritectic and eutectic phase 
diagrams, invariant regions are shown by thin horizontal lines). 
Stable solid domain at low $x_2$ splits into two.
One of them ($\alpha$) corresponds to an almost pure crystal made of the
lighter element. It is characterized by gradually decreasing 
maximum $x_2$ and it survives till much higher $Z_2$, including 
C/Mg mixture at $Z_2=12$. The other ($\gamma$) is short-lived. 
It is represented by a diminishing island, which is 
last seen at $Z_2 \sim 10.3$. 

Starting from $Z_2 \sim 10.5$, the
diagram assumes a chrestomathic eutectic shape. Near the eutectic point,
the right-hand section of the liquidus demonstrates a tendency to 
steepening, which, by $Z_2/Z_1=3$, results in a breaking of the 
liquidus and transformation to a diagram with a liquid gap 
(see section \ref{HeUsec}).   

If $S_{\rm mixZ} \to S_{\rm mix}$ in equation 
(\ref{Fsol}), the sequence of
phase diagrams remains nearly the same as in Fig.\ \ref{shapes}, but 
the respective shapes occur at $Z_2$, which are $\sim 0.1$ smaller 
than in Fig.\ \ref{shapes}. In particular, there are two separate 
stable solid domains at low $x_2$ for C/Ne mixture with $S_{\rm mix}$,
looking very similar to $Z_2=10.1$ graph in Fig.\ \ref{shapes} 
(see Fig.\ \ref{cne_diag}a).

\subsection{$^{12}$C/$^{20}$Ne and $^{12}$C/$^{24}$Mg mixtures}
\label{UNeUMg}
\begin{figure*}
\begin{center}
\leavevmode
\includegraphics[bb=71 558 568 741, width=170mm]{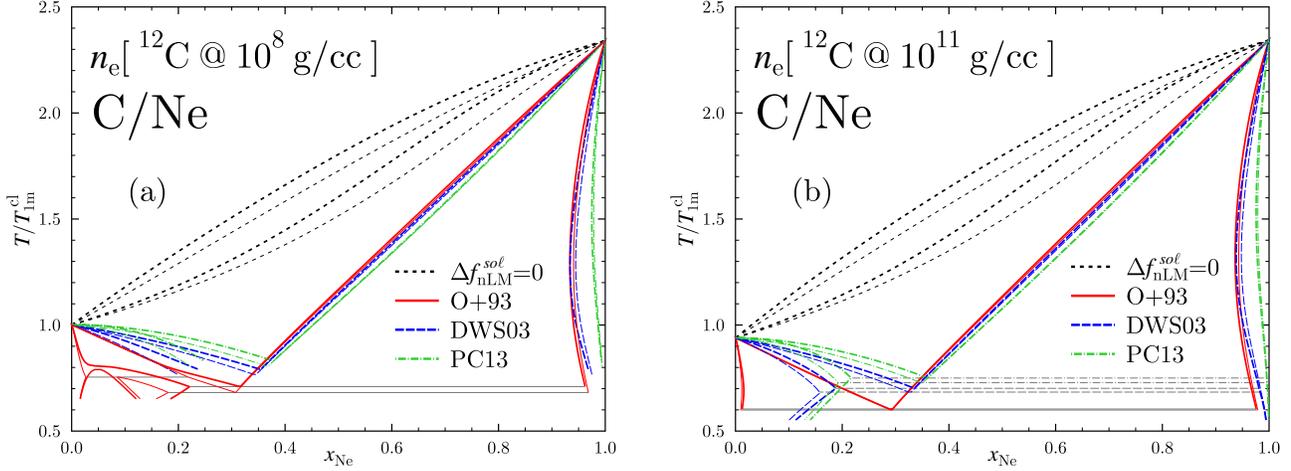}
\end{center}
\vspace{-0.4cm}
\caption[ ]{Same as in Fig.\ \ref{co_diag} but for C/Ne mixture. Grey
horizontal lines display eutectic invariant regions. For clarity, 
dashed and dot-dashed curves are not shown in panel (a) below their
eutectic points. They would look exactly the same as in panel (b).}
\label{cne_diag}
\end{figure*}

The results for carbon/neon mixture are displayed in more detail in 
Fig.\ \ref{cne_diag}. Electron densities, type, and colour of the 
curves are the same as in Fig.\ \ref{co_diag}, so that panel 
(a) corresponds to classic, while panel (b) to moderately quantum 
densities. Grey horizontal lines display eutectic invariant
regions for various $\Delta f^{{\sl so}\ell}_{\rm nLM}$ 
parameterizations and residual entropies. To avoid cluttering, we do 
not extend dashed and dot-dashed curves below their eutectic points 
in panel (a) and do not show respective invariant regions. These 
features would look exactly the same as in panel (b).

In section \ref{varyZ2}, we have already discussed the complex 
low-$x_{\rm Ne}$ corner of the phase diagram for the \citet{O+93} fit 
in panel (a). Remarkably, the ``complexity'' disappears in 
panel (b). As we have mentioned, the solid 
solubility curves are not affected by ion quantum effects. Hence, the 
reason for the change is a ``consumption'' of the two-solid region by 
the liquid-solid structure moving to lower temperatures due to ion 
quantization. As a result, 
for $x_{\rm Ne} \lesssim 0.3$ in panel (b), the free energy minimum is 
realized by a solid solution with an extremely low neon fraction, a 
liquid solution, or a mechanical mixture of the two. The thick 
solid (red) curve in ``quantum'' panel (b) thus achieves a more 
advanced degree of transformation to the eutectic type, looking similar 
to the classic mixture diagram at $Z_2=10.5$ in Fig.\ \ref{shapes}.

An important conclusion is that, for some charge ratios, rather mild ion 
quantum effects, considered in this work, may result in a qualitative
modification of the phase diagram. Same goes for the residual
entropy choice as illustrated by thick and thin solid (red) curves in 
Fig.\ \ref{cne_diag}a. In particular, quantum effects are expected to 
modify the low-$x_{\rm Ne}$ and low-$x_{\rm O}$ corner of the ternary 
C/O/Ne phase diagram at high densities \citep[cf. Fig.\ 4a of][]{CHC20}. 
This may be relevant for the outcome of $^{22}$Ne distillation 
invoked for an explanation of observed cooling delays of ultramassive
WD \citep[][]{BDS21}.  

The other $\Delta f^{{\sl so}\ell}_{\rm nLM}$ parameterizations 
produce quantitatively similar to each other, standard looking 
eutectic phase 
diagrams, in which neon fraction in a carbon crystal can exceed 20\%
but carbon fraction in neon crystals is quite low \citep[especially 
for the fit of][]{PC13}. The only quantum effect that can be 
seen for these $\Delta f^{{\sl so}\ell}_{\rm nLM}$ expressions is a 
reduction of the pure carbon melting temperature and an associated 
reduction of the eutectic temperatures.
 
\begin{figure*}
\begin{center}
\leavevmode
\includegraphics[bb=72 553 569 740, width=170mm]{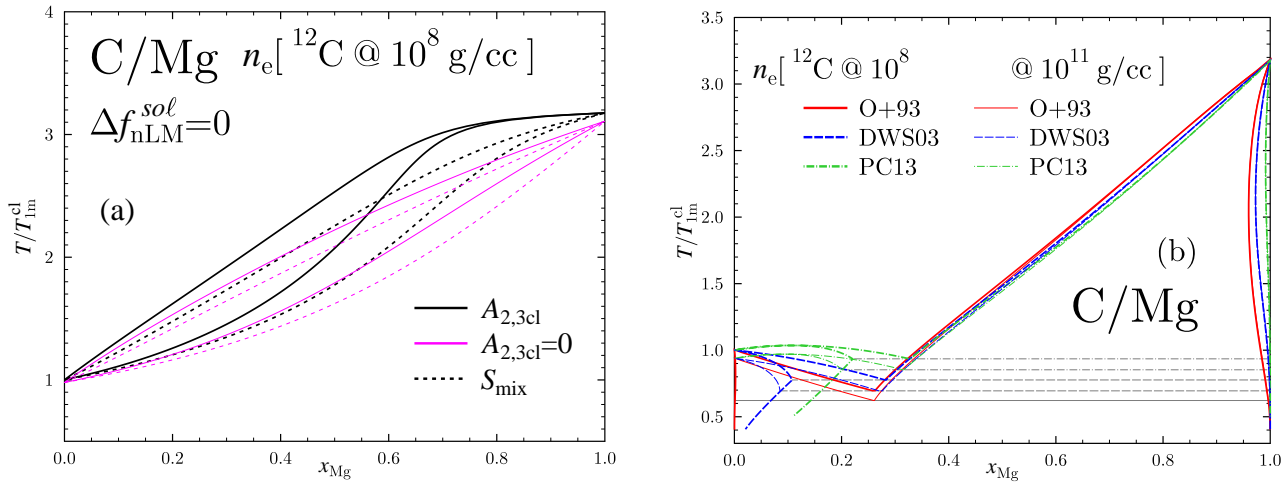}
\end{center}
\vspace{-0.4cm}
\caption[ ]{Phase diagram for C/Mg mixture. (a) Classic mixture with
$\Delta f^{{\sl so}\ell}_{\rm nLM} = 0$. Solid thick (black) and thin 
(magenta) curves display calculations with full $A_{\rm 2,3cl}$ in the
solid phase and with $A_{\rm 2,3cl}=0$, respectively. Dashed curves
are obtained from the solid ones by replacement 
$S_{\rm mixZ} \to S_{\rm mix}$. (b) Thick and thin curves correspond
to classic and quantum densities, respectively, with 
$\Delta f^{{\sl so}\ell}_{\rm nLM}$ given by fits of \citet{O+93} 
(solid, red), \citet{DWS03} (dashed, blue), and \citet{PC13} 
(dot-dashed, green). Horizontal lines are invariant regions.}
\label{cmg_diag}
\end{figure*}

Before proceeding further, let us make a technical 
note. In Fig.\ \ref{cne_diag}, 
for $\Delta f^{{\sl so}\ell}_{\rm nLM} = 0$ phase diagram, there is a 
noticeable distortion of the spindle near its upper right corner. 
It is associated with higher order crystal anharmonic corrections 
employed at rather low $\Gamma < \Gamma_{\rm m}$. We try to elucidate 
this effect in Fig.\ \ref{cmg_diag}a for carbon/magnesium mixture. In 
this case, thick (black) and thin (magenta) solid curves display phase 
diagrams for $\Delta f^{{\sl so}\ell}_{\rm nLM}=0$ assuming full 
equations for crystal thermodynamics \citep[with classic anharmonic 
coefficients $A_{\rm 2,3cl}$ as reported in][]{BC22} and assuming
$A_{\rm 2,3cl}=0$ \citep[as suggested earlier by][]{MC10}, respectively. 
Nulling $A_{\rm 2,3cl}$ also slightly affects melting 
points for both pure carbon and pure magnesium crystals. Dashed curves 
assume $S_{\rm mix}$ for the residual entropy. For $Z_2/Z_1$ 
considered so far, we have been checking the robustness of our 
conclusions against setting $A_{\rm 2,3cl}$ to zero and have observed 
no issues. 

In Fig.\ \ref{cmg_diag}b, we show C/Mg phase diagrams for 
$\Delta f^{{\sl so}\ell}_{\rm nLM}\ne 0$. Thick curves correspond to
a classic system, whereas thin ones illustrate ion quantum effects.
The latter are seen to be mild. They do not produce 
qualitative changes for this charge ratio. We kept
$A_{\rm 2,3cl}$ to their original values; setting them to 0 
results in a slight, almost uniform shift of the phase diagrams down 
related to the reduction of melting temperatures for both pure 
crystals. The general shape of the phase diagram and
the effect of different $\Delta f^{{\sl so}\ell}_{\rm nLM}$ 
parameterizations are virtually the same as for already discussed 
quantum C/Ne mixture.        

Summarizing, in contrast to systems with a smaller charge ratio, 
C/Ne and C/Mg mixtures, for $\Delta f^{{\sl so}\ell}_{\rm nLM} \ne 0$, 
exhibit phase diagrams of the eutectic type. These contain two widely 
separated  solid domains, enriched in lighter or heavier ion species.
At $x_{\rm Ne} \sim 0.25-0.35$, there is a eutectic point, below 
which liquid does not exist. Cooling in eutectic systems is known to 
be accompanied by interesting nonequilibrium phenomena \citep[e.g. 
section 6 of][]{G68}. In general, depending on the initial composition 
of a cooling liquid, the liquidus can be reached to the left or to the 
right of the eutectic point. This will cause precipitation of, 
respectively, $Z_1$- or $Z_2$-rich solids called primary 
solidification. Further cooling will proceed along the liquidus. If 
the initial composition is such that the eutectic point can be reached 
during this process, the liquid will completely freeze, separating 
into $Z_1$- and $Z_2$-rich crystallites. However, through the effect 
known as constitutional undercooling, a quasi-periodic lamellar 
superstructure may actually form. This, along with the primary grains, 
affects the overall crystal strength and disorder and may be important 
for elastic, kinetic, and other properties of these layers.

\subsection{$^4$He/$^{12}$C mixture}
\label{HeUsec}
Finally, let us briefly describe a phase diagram for a mixture with a 
rather large $Z_2/Z_1=3$. We shall analyze this case using the 
example of helium/carbon mixture, which is relevant for accreting WD. 
The results should 
be indistinguishable if any other $Z_1$ is used, provided the 
densities are rescaled appropriately to achieve the same strength of ion
quantum effects. For this charge ratio, 
the $\Delta f^{{\sl so}\ell}_{\rm nLM}$ fit  
by \citet{DWS03} is beyond its applicability limits. 
Hence, we only consider
parameterizations by \citet{O+93} and \citet{PC13}, assuming 
$A_{\rm 2,3cl}=0$ (cf.\ section \ref{UNeUMg}).

\begin{figure}
\begin{center}
\leavevmode
\includegraphics[bb=200 178 440 741, width=84mm]{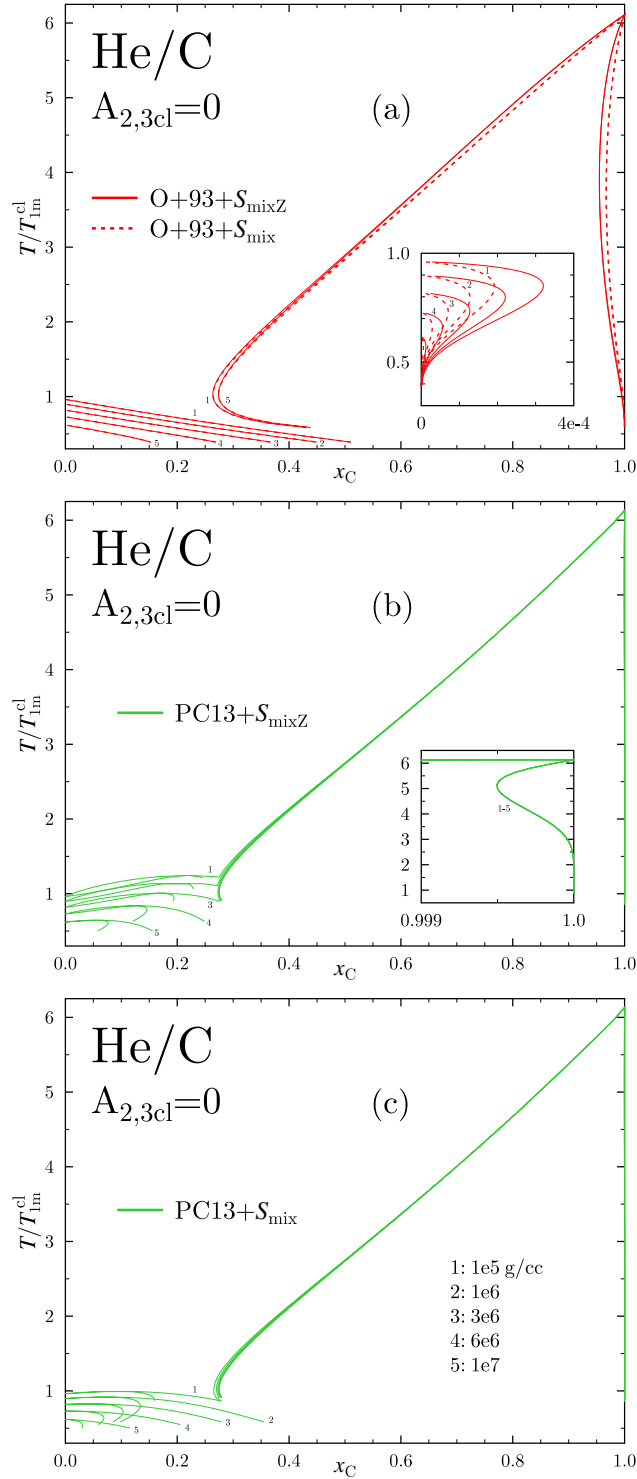}
\end{center}
\vspace{-0.4cm}
\caption[ ]{He/C phase diagram for $\Delta f^{{\sl so}\ell}_{\rm nLM}$
given by \citet{O+93} (panel a); \citet{PC13} with $S_{\rm mixZ}$ 
(panel b); \citet{PC13} with $S_{\rm mix}$ (panel c). Solid and dashed 
curves in panel (a) correspond to calculations with $S_{\rm mixZ}$ 
and $S_{\rm mix}$, respectively. Curves marked by numerals 
1--5 illustrate ion quantum effects and correspond to different 
electron densities listed in the legend in panel (c).}
\label{hec}
\end{figure}

In Fig.\ \ref{hec}a, the phase diagram based on the fit by
\citet{O+93} is shown for several electron densities. One observes two 
main structures. First, there is a large unstable ``triangle'' at 
intermediate and large $x_{\rm C}$ reminiscent of eutectic phase 
diagrams at lower $Z_2/Z_1$. On top and to the left of it, 
there is a stable liquid domain. To the right of the triangle, 
there is a narrow stable solid region, where helium fraction must 
stay below 5\%. Solid and dashed curves have the residual solid 
mixture entropy equal to $S_{\rm mixZ}$ and $S_{\rm mix}$, 
respectively. Two solid lines (merging for $x_{\rm C} > 0.45$) 
are plotted for different electron densities. (Same applies to two 
merging dashed lines.) The outer solid (or dashed) line assumes the 
electron density of pure helium at $10^5$ g cm$^{-3}$, the inner 
one $10^7$ g cm$^{-3}$. We have also checked $10^8$ and 
$10^9$ g cm$^{-3}$, which resulted in a further minuscule inward 
motion of these curves.    

At low $x_{\rm C}$, there is another structure which should be 
analyzed together with the curves shown in the inset. 
Five curves from top to bottom represent liquidi for electron 
densities corresponding to pure helium at $10^5$, $10^6$, 
$3 \times 10^6$, $6 \times 10^6$, and $10^7$ g cm$^{-3}$, respectively.
The associated solidi are shown in the inset. One observes an extremely 
narrow stable sector of an almost pure helium crystal with a carbon 
admixture
not exceeding a few times $10^{-4}$. With density increase, the allowed
carbon fraction steadily drops. We have also considered higher electron
densities. The reduction of the structure continued, and it was not seen 
in the scale of Fig.\ \ref{hec}a already at $10^8$ g cm$^{-3}$.

One sees that, at no density, the two structures intersect, no eutectic 
point forms, and a stable liquid solution between the unstable regions 
is predicted to exist down to very low temperatures. 
The liquid gap becomes wider as ion quantum effects become stronger 
(i.e. as the density increases). Of course, the calculations become
less and less reliable with decrease of temperature, because liquid
fits are used well below crystallization.    

In panels (b) and (c) of Fig.\ \ref{hec}, we show phase diagrams based
on $\Delta f^{{\sl so}\ell}_{\rm nLM}$ fit proposed by \citet{PC13}
with the residual entropy equal to $S_{\rm mixZ}$ and $S_{\rm mix}$, 
respectively. The general features of these phase diagrams are similar
to the diagram shown in panel (a). In particular, there is a 
triangular unstable region at $x_{\rm C} \gtrsim 0.3$, whose boundaries
are weakly sensitive to the density. In contrast to panel (a), 
very low fraction of helium ($<0.0005$, see the inset, where all 5 
curves have merged) is predicted in almost pure stable carbon crystals.

In the low-$x_{\rm C}$ region, the diagram in panel (b) has a narrow
spindle-shaped unstable area, below and to the right of which a 
helium crystal with a rather large fraction of carbon (up to 
$\sim 25\%$ at classic densities) is predicted. The maximum carbon 
fraction decreases as the system becomes more quantum. The crystal 
domain is bounded from the right by another unstable zone, whose right 
boundary (liquidus) may intersect liquidus of the 
large-$x_{\rm C}$ unstable region. This leads to a formation of a 
eutectic point and an extensive two-solid unstable region. However, as 
ion quantum effects get stronger, the two structures cease 
intersecting, and a liquid gap between them, similar to the one in 
panel (a), appears.
 
If the residual entropy is equal to $S_{\rm mix}$ (panel c), 
the diagram is very similar to panel (b). However, the 
spindle-shaped unstable zone is degenerated, and the maximum fraction 
of carbon in a helium-dominated crystal does not exceed $\sim 15\%$.

\section{Conclusion}
\label{concl}
We have studied phase diagrams of binary ionic mixtures in the 
$1 < Z_2/Z_1 \leq 3$ range using the linear mixing with corrections 
approach and taking advantage of the recent progress in understanding 
quantum OCP thermodynamics. We have compared three fits for
corrections to linear-mixing energies in the solid phase and two 
different models of solid residual entropy at $T=0$. We have used 
actual fits of OCP thermodynamic functions in the entire required 
$\Gamma$ ranges, including, where needed, solid phase formulae at 
temperatures above melting and vice versa, without resorting to any 
extrapolation schemes. This is the least reliable aspect of BIM 
thermodynamic calculations based on the linear mixing theory. It can 
be improved via first-principle analysis of the mixtures. 
Our results for the C/O phase diagram at classic conditions with 
non-linear-mixing corrections given by \citet{O+93} and with the 
standard choice for the solid residual entropy agree well with 
previous work \citep[][]{MC10,B+20}. 

For solid C/O and O/Ne mixtures, we have discovered extensive 
miscibility gaps. Their appearance seems to be a robust feature of the
theory: the gaps evolve naturally into two-solid regions of eutectic 
phase diagrams predicted at higher $Z_2/Z_1$ and also they do not rely 
on fit extensions beyond applicability limits because both components 
are solid. The gaps are very sensitive to BIM composition and physics
being strongly different for C/O and O/Ne mixtures and for the three
variants of the $\Delta f^{{\sl so}\ell}_{\rm nLM}$ fit. 

When matter cools to its miscibility gap temperature, the exsolution 
process (well-known on Earth) starts taking place. It results in a 
separation of heavier and lighter ($Z_2$- and $Z_1$-rich) solid 
solutions, which may represent a significant reservoir of gravitational 
energy. It would be interesting to include this effect in future
WD cooling simulations. 

At higher $Z_2/Z_1$, our analysis indicates a transformation of 
azeotropic phase diagrams into peritectic and eventually to eutectic
ones. Cooling and freezing in such systems may result in 
a precipitation of $Z_1$- or $Z_2$-rich solids. A nonequilibrium
effect of constitutional undercooling may produce quasi-periodic 
superstructures of various geometries. All these processes affect 
crystal strength, kinetics, and other important properties of matter.

Ion quantum effects, for the most part, result in moderate modifications 
of phase diagrams, which can be traced back to melting temperature 
decrease in quantum one-component systems. However, 
for certain charge ratios, a BIM may be close to a 
qualitative transformation of its phase diagram, in which case ion 
quantum effects may produce drastic restructuring of the latter 
(e.g. Fig.\ \ref{cne_diag}). This is expected to affect also the 
low-$x_{\rm Ne}$ and low-$x_{\rm O}$ corner of the ternary 
C/O/Ne phase diagram at high densities, which may be relevant for the 
outcome of $^{22}$Ne distillation process in ultramassive WD. 

In general, phase diagrams constructed for non-zero corrections to 
solid-phase linear-mixing energies exhibit much stronger separation of 
ionic species upon mixture crystallization than those in 
the model without corrections, especially at $Z_2/Z_1 \gtrsim 1.5$.  
The separation obtained from fits of \citet{DWS03} and \citet{PC13} is 
noticeably greater than that stemming from the parameterization by 
\citet{O+93} for smaller charge ratios (such as O/Ne or C/O mixtures). 
For eutectic systems, these two groups of fits differ quite a bit in 
terms of stable solid solution compositions and eutectic point 
positions. The question of which of these results is closer to the 
reality can be settled via first-principle analysis. The most recent 
classic GDMC study of \citet{B+20} for C/O mixture indicates that the
fit of \citet{O+93} is likely closer to the truth at this charge 
ratio.
  
We have found only rather weak sensitivity to the residual entropy 
choice across the parameter space. For C/O mixture, replacing 
$S_{\rm mixZ} \to S_{\rm mix}$ worsens the agreement with the curves
reported by \citet{B+20} (Fig.\ \ref{co_compar}).  

We anticipate our results to enable more reliable modelling of 
ion redistribution at melting/crystallization and after it, 
gravitational energy release, thermodynamics, kinetics, elasticity etc 
in a variety of astrophysical phenomena associated with compact stars.

\section*{Acknowledgments}
This work, except for subsection \ref{astro}, was supported by RSF 
grant 19-12-00133-P. The author is grateful to A. I. Chugunov for 
discussions at the initial stages of this work and for useful critique 
of the manuscript. The author is deeply thankful to anonymous referee 
for valuable remarks.

\section*{Data Availability}
The data underlying this article will be shared on reasonable 
request to the author.

\end{document}